\documentclass[prl,nofootinbib,twocolumn]{revtex4}
\def\mysection#1{{\bf #1.} }
\def\mysections#1{{\bf #1.} }

\usepackage{amssymb}
\usepackage{amsmath}
\usepackage[dvips]{graphicx}
\usepackage{longtable}
\usepackage{verbatim}
\usepackage{amsfonts}
\usepackage{subfigure}

\newcommand{\avg}[1]{\left\langle {#1} \right\rangle}
\newcommand{\abs}[1]{\lvert#1\rvert}

\newcommand{\be}{\begin{equation}}
\newcommand{\ee}{\end{equation}}
\newcommand{\bea}{\begin{eqnarray}}
\newcommand{\eea}{\end{eqnarray}}
\newcommand{\beq}{\begin{equation}}
\newcommand{\eeq}{\end{equation}}
\def\beqa{\begin{eqnarray}}
  \def\eeqa{\end{eqnarray}}
\newcommand{\bv}{\left(\begin{array}{c}}
\newcommand{\ev}{\end{array}\right)}
\newcommand{\no}{\nonumber}
\def\lsim{\mathrel{\rlap{\lower4pt\hbox{\hskip1pt$\sim$}}
    \raise1pt\hbox{$<$}}}         %less than or approx. symbol
\def\gsim{\mathrel{\rlap{\lower4pt\hbox{\hskip1pt$\sim$}}
    \raise1pt\hbox{$>$}}}         %greater than or approx. symbol

\begin{document}

\vspace*{-30mm}

\title{A tale of two Higgs}

\author{Aielet Efrati}\email{aielet.efrati@weizmann.ac.il}
\affiliation{Department of Particle Physics and Astrophysics,
  Weizmann Institute of Science, Rehovot 76100, Israel}

\author{Daniel Grossman}\email{daniel.grossman@weizmann.ac.il}
\affiliation{Department of Particle Physics and Astrophysics,
  Weizmann Institute of Science, Rehovot 76100, Israel}

\author{Yonit Hochberg}\email{yonit.hochberg@weizmann.ac.il}
\affiliation{Department of Particle Physics and Astrophysics,
  Weizmann Institute of Science, Rehovot 76100, Israel}

\vspace*{1cm}

\begin{abstract}
A new boson with mass $\sim$125~GeV and properties similar to the Standard Model Higgs has been discovered by both the ATLAS and CMS collaborations, with significant observation in the $ZZ^*\to 4\ell$ and $\gamma\gamma$ channels. In this work we ask whether the signals in these two channels can be due primarily to two distinct resonances, each contributing dominantly to one channel. We investigate this question in the framework of a 2HDM and several of its extensions. We conservatively find that such a scenario is not possible in a pure 2HDM, nor under the addition of vector-like quarks, but is allowed when adding one or two top-like scalars, if one allows for sub-one $\tan\beta$.
The resonances in the diboson and diphoton channels can then be two scalars, or a scalar and a pseudoscalar, respectively. In each viable case, we further find the expected future deviations in the diboson, diphoton, $b\bar b$ and $\tau\tau$ rates, which will be useful in excluding the two-resonance scenario.
\end{abstract}

\maketitle

%%%%%%%%%%%%%%%%%%%%
\mysection{Introduction}
The ATLAS~\cite{:2012gk} and CMS~\cite{:2012gu} collaborations have recently announced the discovery of a new boson with mass in the vicinity of $125$~GeV and properties resembling that of a Standard Model (SM) Higgs. The cleanest search channels for the Higgs are the $h\to \gamma\gamma$ and $h\to ZZ^*\to 4\ell$ modes, where until very recently good agreement between the mass of the resonance in each channel was observed. The latest ATLAS data for these two channels exhibits a mass discrepancy~\cite{ATLASdis}, and though it is likely a statistical feature that may soon disappear, it is interesting to ask the following question: Can the resonance observed in the diphoton channel be different from that observed in the $ZZ^*$ channel? Namely, can the two signals observed in these modes be due primarily to two separate particles that are close in mass? This is the question we aim to address in this work, within the context of a 2~Higgs Doublet Model (2HDM). Though inspired by the current trend in the ATLAS data, we do not attempt to explain, reproduce or fit to the full Higgs data set under the assumption of a two particle hypothesis. Rather, we are interested in addressing the qualitative question of whether the Higgs data could be due to two resonances; a question which is relevant even if the current mass discrepancy disappears.

%%%%%%%%%%%%%%%%%%%%
\mysection{Preliminaries}\label{sec:pre}
We start by setting the notation to be used along this paper. Assuming two distinct particles are observed in the latest LHC results, we denote the particles observed in the $ZZ^*$ and the diphoton channels as $\phi^1$ and $\phi^2$, respectively. We define the normalized production rate times branching ratio of a given channel by
\beq\label{eq:ratios}
R_X^i = \frac{\sigma_{\rm tot}^i\times {\rm BR} \left( \phi^i \rightarrow X \right)  }{\sigma_{\rm tot}^{\rm SM}\times {\rm BR}^{\rm SM} \left( h \rightarrow X \right) }\,,
\eeq
where $X=ZZ^*,\gamma\gamma$, $i = 1,2$ for the different scalars, and $\sigma_{\rm tot}^{\rm SM}$ and ${\rm BR}^{\rm SM}$ are the production cross section and branching ratios of a 125~GeV SM Higgs boson. Assuming the mass splitting between the two resonances is larger than their width, interference effects are suppressed, and the total production rate times branching ratio in each channel is the sum of the two contributions, $R_X = R_X^1 + R_X^2$.

We ask whether the resonances observed in the different channels can be a result of two different particles of mass close to~$125$~GeV. For this to be the case, the following conditions should hold:
\begin{enumerate}
\item In each channel a different particle is responsible for the major contribution to the production rate times branching ratio, while the contribution of the other particle is minor;
\item Together, the two resonances give the observed rates for the two channels.
\end{enumerate}
The rates in additional measured channels, such as the $b\bar b$ and $\tau\tau$ modes, should also be accommodated by the two particles combined.

We quantify these conditions by the following (mild) set of requirements:
\beqa\label{eq:cond}
R_{ZZ^*}^1 &\geq& 3 R_{ZZ^*}^2 \,, \no \\
R_{\gamma \gamma}^2 &\geq& 3 R_{\gamma \gamma}^1 \,, \no \\
0.5 \leq &R_{ZZ^*}& \leq 2.5 \,, \no \\
0.5 \leq &R_{\gamma \gamma}& \leq 2.5\,.
\eeqa
We (conservatively) consider a parameter region to be disfavored if it fails any of the four conditions of Eq.~\eqref{eq:cond}, and take a model to be ruled out if there is no point in its parameter space that can satisfy all four of these conditions. In what follows, we refer to the first and last two lines of Eq.~\eqref{eq:cond} as the `ratio' and `range' conditions, respectively. The use of such a moderate set of requirements allows us to be conservative: A model that fails in accommodating the range conditions cannot be the source of the Higgs measurements in the $ZZ^*$ and diphoton channels, regardless of the ratio conditions. A model that fails the ratio conditions cannot explain the Higgs measurements in the $ZZ^*$ and diphoton channels by means of two separate particles. A model that passes the criteria of Eq.~\eqref{eq:cond} can then be scrutinized in order to learn whether the various rates are enhanced, suppressed or similar to that of the SM. We further restrict to similar range conditions in the $b\bar b$ and $\tau\tau$ channels, requiring that a model that passes the conditions of Eq.~\eqref{eq:cond} also obeys
\beq\label{eq:condb}
0.5 \leq R_{b\bar b}\leq 2.5\,,\ \ \
0.5 \leq R_{\tau\tau}\leq 2.5\,,
\eeq
in order to be viable.

A comment is in order regarding existing literature on the presence of multiple resonances in the Higgs data. This topic was discussed in~\cite{Gunion:2012gc} in the context of the NMSSM, in~\cite{Ferreira:2012nv} in the context of a 2HDM,
and most recently in~\cite{Grossman:2013pt} where general conditions for detection of such resonances were derived (see also~\cite{Gunion:2012he}). However, these works do not require the signals in each channel to be predominantly a result of a distinct particle and thus address a different qualitative question than that posed in this work.

We follow the notations of~\cite{Carmi:2012in}, and write the Lagrangian of the two scalars $\left( i = 1 , 2\right)$ as follows:
\beqa\label{eq:Leff}
\mathcal{L}_{\phi^i} & \supset &
c_W^i \frac{2 m_W^2}{v} \phi^i W^+_\mu W^-_\mu
+ c_Z^i \frac{m_Z^2}{v} \phi^i Z_\mu Z_\mu
- c_t^i \frac{m_t}{v} \phi^i \bar{t} t \no \\
& - & c_b^i \frac{m_b}{v} \phi^i \bar{b} b
- c_\tau^i \frac{m_\tau}{v} \phi^i \bar{\tau} \tau
- c_c^i \frac{m_c}{v} \phi^i \bar{c} c \no \\
& - & c_f^i \frac{m_f}{v} \phi^i \bar{f} f
- c^i_s \frac{2 m_s^2 }{v} \phi^i S^\dagger S\,,
\eeqa
where custodial symmetry implies $c_W^i = c_Z^i \equiv c_V^i$, and we assume the invisible branching ratios to be negligible. $f \left( S\right)$ symbols any new fermion (scalar) added to the theory, and we do not consider the addition of vector bosons. The effective couplings to gluons and photons are obtained via one loop processes, defined such that
\beqa
\Gamma_{gg}^i=\frac{|\hat c_{g}^i|^2}{|\hat c_{g}^{\rm SM}|^2}\Gamma_{gg}^{\rm SM}\,,\ \ \
\Gamma_{\gamma\gamma}^i=\frac{|\hat c_{\gamma}^i|^2}{|\hat c_{\gamma}^{\rm SM}|^2}\Gamma_{\gamma\gamma}^{\rm SM}\,,
\eeqa
where $\Gamma_X^i$ denotes the partial width into $X$, and are given by:
\beqa
\hat{c}_g^i &=& c^i_t A_f (\tau_t^i) + c^i_b A_f ( \tau_b^i) + c^i_c A_f ( \tau_c^i)  \no \\
& + & 2 C(r_f) c^i_f A_f(\tau_f^i) + \frac{ C (r_s) }{ 2 } c^i_s A_s ( \tau_s^i ) \,, \no \\
\hat{c}_\gamma^i &=& \frac{2}{9} c^i_t A_f ( \tau_t^i ) - \frac{7}{8} c^i_V A_f ( \tau_W^i ) + \frac{1}{18} c^i_b A_f ( \tau_b^i ) + \frac{1}{6} c^i_\tau A_f ( \tau_\tau^i ) \no \\
& + & \frac{N ( r_f )}{ 6 } Q_f^2 c^i_f A_f ( \tau_f^i ) + \frac{N ( r_s )}{ 24 } Q_s^2 c^i_s A_s ( \tau_s^i )\,.
\eeqa
Here ${\rm Tr} \left( T^a T^b \right)=C \left(r\right) \delta^{ab}$, $N (r)$ is the dimension of the representation $r$, $\tau_x^i = m_{\phi^i}^2 / 4 m_x^2 $ and $A_{f,s}$ are the fermion and scalar loop functions given by~\cite{Carmi:2012in,Gunion:1989we}:
\beqa
A_f \left( \tau \right) &=& \frac{\xi + 2 }{2 \tau^2} \left[ \xi \tau + \left( \tau - \xi \right) f \left( \tau \right) \right] \, ,\no\\
A_s \left( \tau \right) &=& \frac{3}{\tau^2} \left[ f\left( \tau \right) - \tau \right] \, ,
\eeqa
with $\xi = 1 \, (0)$ for scalars (pseudoscalars) and
\beq
f(\tau)=\left\{\begin{array}{ll}[\sin^{-1}(\sqrt{1/\tau})]^2, & {\rm if}\ \tau\geq1,\\ -\frac{1}{4}[\ln (\eta_+/\eta_-)-i\pi]^2, & {\rm if} \ \tau<1\,.\end{array}\right.
\eeq
For heavy particles, $\tau_x^i \ll 1 $, and one finds $A_x \simeq 1$. In the SM with a $125$~GeV Higgs, $c_{t,{\rm SM}}=c_{b,{\rm SM}}=c_{c,{\rm SM}}=c_{\tau,{\rm SM}}=c_{V,{\rm SM}}=1$, $\hat{c}_{g,{\rm SM}}\simeq 1.0$ and $\hat c_{\gamma,{\rm SM}}\simeq-0.82$. Defining the total width as
\beqa
\Gamma_{\rm tot}^i &=& \abs{C_{\rm tot}^i}^2 \Gamma_{\rm tot}^{\rm SM} \,,
\eeqa
with~\cite{Dittmaier:2011ti}
\beqa\label{eq:ctot}
\abs{C_{\rm tot}^i }^2 &\simeq& 0.58 \abs{c^i_b }^2  + 0.24 \abs{c^i_V}^2 + 0.09 \frac{\abs{\hat{c}^i_g}^2}{\abs{\hat{c}_g^{\rm SM} }^2} \no \\
& + & 0.06 \abs{c^i_\tau}^2 + 0.03 \abs{c^i_c}^2\,,
\eeqa
one finds the following approximate expressions:
\beqa
R^i_{ZZ^*} \simeq \left|\frac{\hat{ c}^i_g c^i_V }{\hat{ c}_g^{\rm SM} C^i_{\rm tot} }\right|^2 \,, \no \\
R^i_{\gamma \gamma} \simeq \left|\frac{\hat{ c}^i_g \hat{ c}^i_\gamma  }{\hat{ c}_g^{\rm SM} \hat{c}_\gamma^{\rm SM} C^i_{\rm tot} }\right|^2\,,
\eeqa
where in the above we have assumed that the production is dominated by the gluon fusion process. Expressions for the $b\bar b$ rate with an associated vector boson, $R_{b\bar b}$, as well as the $\tau\tau$ rate $R_{\tau\tau}$, and the dijet category in the diphoton channel $R_{\gamma\gamma}^{\rm jj}$, can be found in or deduced from~\cite{Carmi:2012in}, with the efficiencies in the diphoton channel, both inclusive and dijet categories, taken from~\cite{ATLASdiphoton}. In the $\tau\tau$ channel, we consider the vector boson fusion category, and use the efficiency numbers of the semi-leptonic channel as given in~\cite{ATLAStau}.

In what follows we use the full expressions for the production cross section times branching ratios, Eq.~\eqref{eq:ratios}, and ask whether the conditions detailed in Eq.~\eqref{eq:cond} can be satisfied by two bosons within the 2HDM framework. We consider four different scenarios: Pure 2HDM without additional particles,
2HDM plus vector-like quarks in the $\left( 3,1 \right)_{2/3}$ and $\left( 3,1 \right)_{-1/3}$ representations, and 2HDM plus additional one or two top-like scalars.

Throughout this paper we use the leading order contributions to the relevant production and decay processes. In all models considered, the QCD next-to-leading-order (NLO) corrections are either small or factored out (see {\it e.g.}~\cite{Spira:1995rr,Djouadi:1999ht}), and thus cancel in the ratios to the SM. The electroweak NLO corrections are small in the discussed models (see {\it e.g.}~\cite{Djouadi:1997rj}). In this context, a comment is in order regarding addition of heavy chiral fermions which are not included in this work. Since chiral fermions acquire mass from electroweak symmetry breaking (EWSB), their loop contributions to the relevant production and decay channels do not decouple as their masses increase. The electroweak NLO contributions can then induce ${\cal O}(1)$ corrections to all decay modes~\cite{Djouadi:1997rj}, and are important. The proper treatment of additional heavy chiral fermions is beyond the scope of this work, and is hence omitted from further discussion.

%%%%%%%%%%%%%%%%%%%%
\mysection{2HDM}\label{sec:2HDM}
We now study the pure CP-conserving 2HDM scheme with no additional particles, considering both type~I and type~II couplings. The dependence of the results on the lepton couplings in the pure 2HDM is highly suppressed, allowing us to extend our conclusions to the other natural flavor conserving 2HDM types. We conventionally define the ratio between the VEVs of the two Higgs doublets by $\tan \beta$ and the rotation angle to the neutral mass eigenstates by $\alpha$. The neutral spectrum of the 2HDM contains two scalars $h$ and $H$ with $m_h<m_H$, and one pseudoscalar $A$.

The couplings of $h,H$ and $A$ to vector bosons, up quarks, down quarks and charged leptons are presented in Table~\ref{tab:couplings}. We define an effective coupling between the pseudoscalar and two massive gauge bosons, $|c_{AVV}^{i}|^2 \equiv \Gamma_{A \rightarrow VV}^i / \Gamma_{h \rightarrow VV}^{\rm SM} $, where $\Gamma_{A \rightarrow VV}^i $ is the appropriate partial width of the pseudoscalar in type $i=$I,~II~\cite{Djouadi:2005gi,Djouadi:2005gj,Branco:2011iw}, arising from the dimension 5 operator $A V_{\mu \nu} \tilde{V}^{\mu \nu}$ which is generated at the loop level, where $V_{\mu\nu}$ is the field strength of the vector boson. Throughout we neglect the contribution of the charged Higgs loops to the diphoton channel (see {\it e.g.}~\cite{Boudjema:2001ii}). We have verified numerically that subject to electroweak precision data~\cite{PDG} and perturbativity constraints on the trilinear couplings, the inclusion of such effects does not alter our results.

\begin{table}[t]
\caption{The dependence on $\alpha, \beta$ of the couplings of the neutral scalars to different particles~\cite{Gunion:1989we}.}
\label{tab:couplings}
\begin{center}
\begin{tabular}{|c|c|c|c|} \hline
\rule{0pt}{1.2em}%
Type I & $\phi VV$\ &  $\phi \bar{u}u$ & $\phi \bar{d}d$ and $\phi \ell^+\ell^-$ \cr \hline
$h$ & $\sin(\beta-\alpha)$ & $\frac{\cos\alpha}{\sin\beta}$\ & $\frac{\cos\alpha}{\sin\beta}$ \cr
$H$ & $\cos(\beta-\alpha)$ & $\frac{\sin\alpha}{\sin\beta}$\ & $\frac{\sin\alpha}{\sin\beta}$ \cr
$A$ & $c_{AVV}^{\rm I}$ & $\cot\beta$ & $- \cot \beta$\cr
\hline\hline
Type II & $\phi VV$\ &  $\phi \bar{u}u$ & $\phi \bar{d}d$ and $\phi \ell^+\ell^-$ \cr \hline
$h$ & $\sin(\beta-\alpha)$ & $\frac{\cos\alpha}{\sin\beta}$\ & $-\frac{\sin\alpha}{\cos\beta}$ \cr
$H$ & $\cos(\beta-\alpha)$ & $\frac{\sin\alpha}{\sin\beta}$\ & $\frac{\cos\alpha}{\cos\beta}$ \cr
$A$ & $c_{AVV}^{\rm II}$ & $\cot\beta$ & $ \tan\beta$\cr
\hline
\end{tabular}
\end{center}
\end{table}

In the following we consider all possible pairs amongst $h,H$ and $A$ as the potential resonances. The exchange of the two scalars produces almost identical results under a shift of $\alpha\to \alpha\pm \pi/2$. Additionally, while the parity of the observed particle in the $ZZ^*$ channel is currently under scrutiny~\cite{:2012br}, we have included the possibility that $\phi^1$ is the pseudoscalar $A$. In all the explored scenarios we find however that the pseudoscalar cannot successfully serve as the resonance in $ZZ^*$ channel due to its suppressed coupling. We thus quote our results only for the cases of $\left( \phi^1, \phi^2 \right)= \left( h,H\right)$ and $\left( h,A\right)$.

Given that $\phi^1$ is the resonance seen in the $ZZ^*$ channel, the data suggests that this particle should have, to some extent, SM-like couplings, namely $c_V^1$ should be of order one. In contrast, the $\phi^1$ contribution to the diphoton channel is required to be small. The effective coupling to two photons arises at one loop, primarily via internal $W$ and $t$ contributions which interfere destructively with each other, $\hat c_\gamma^i\propto c_t^i-\frac{9}{2}c_V^i$. Thus, it is difficult to find a parameter space in which $\phi^1$ dominates the $ZZ^*$ channel but has only a moderate contribution to the $\gamma \gamma$ final state, unless a delicate cancelation occurs. The complementary demand that $\phi^2$ contributes significantly only to the $\gamma \gamma$ channel is easily achieved if $\phi^2$ is the pseudoscalar $A$, since its coupling to two heavy gauge bosons is loop-induced and thus highly suppressed.

We learn that the four conditions of Eq.~\eqref{eq:cond} are difficult to accommodate generically in a 2HDM. In what follows we identify the scenarios and regions of parameter space in which the appropriate cancelations do occur and the conditions are met. We note that a combination of several effects determines whether a scenario or parameter space is viable or not. Whenever possible, we explain these effects.

We consider $0.5 \leq \tan \beta \leq 65$ and $\abs{ \alpha } \leq \pi / 2$, and first ask whether a parameter space exists in which the observed resonances are a result of two different particles, namely the ratio conditions in Eq.~\eqref{eq:cond} are satisfied. We then further seek the parameter space in which all the conditions are met. Our results are as follows:

\begin{itemize}
\item
The two scalars, $h$ and $H$, cannot provide the observed signals. In type II the ratio conditions can be achieved only for $\tan \beta \lsim 0.6$, where the cancelation between the $W$ and the $t$ loops is increased due to the enhanced top coupling. However, this parameter range typically leads to a very small diphoton rate, since a large cancelation occurs for both of the scalars. In type I, since $c^{h,H}_t = c^{h,H}_b $ there is less freedom for $\alpha$ and $\beta$ and the required cancelation cannot occur, yielding null results for the ratio conditions in the case of the two scalars.

\item
For $\phi^1=h$ and $\phi^2= A$ in the type II 2HDM, all four conditions of Eq.~\eqref{eq:cond} are satisfied for $\tan \beta \lsim 0.6$, in which case $R_{\gamma \gamma}\lsim 1.2$. The reason is that for such small $\tan\beta$, the coupling of the scalar to the top is enhanced and suppresses $R^1_{\gamma \gamma}$, while $R^2_{Z Z^*}$ is naturally small. The pseudoscalar contribution to the diphoton signal, $R^2_{\gamma \gamma} $, is further enhanced at such small $\tan\beta$. However, in the relevant parameter space, the normalized production rate times branching ratio in the $b\bar b$ channel, $R_{b\bar b}$, is suppressed to the level of a few percent, disfavoring this scenario. In type I these particles can obey the desired ratios, but the relevant parameter space yields too small a signal in both channels.
\end{itemize}

We learn that a pure 2HDM of type II can account for the conditions of Eq.~\eqref{eq:cond} if $\phi^1=h$ and $\phi^2=A$, for $\tan\beta\lsim 0.6$, but cannot accommodate a reasonable rate in the $b\bar b$ mode, Eq.~\eqref{eq:condb}, disfavoring the pure 2HDM scenario. We note that in 2HDM, a bound exists on the rate of the associated production of $b\bar b$, stemming from Eq.~\eqref{eq:ctot} and Table~\ref{tab:couplings}, of $R_{b\bar b}\lsim 1.7$.

%%%%%%%%%%%
\mysection{2HDM plus heavy vector-like quarks}
Next we consider the addition of two vector-like top quarks, transforming as $ T \sim \left( 3,1 \right)_{2/3}$ and $T^c \sim \left( \bar{3} ,1 \right)_{- 2/3 } $. Assuming small mixing between these heavy fields and the first two generations, the most general renormalizable interactions are:
\beq\label{eq:Lvq}
\mathcal{L} = \mathcal{L}_{\rm 2HDM} - M T T^c - \lambda Q_3 h_u T^c - m T t^c + {\rm h.c.},
\eeq
where the last term in Eq.~\eqref{eq:Lvq} can be rotated away by a redefinition of $T$ and $t$ without loss of generality, and $h_u$ is the up-type Higgs of the 2HDM with $\avg{h_u} = v_u / \sqrt{2}$. After EWSB the heavy fields mix with the left- and right-handed third generation SM quarks $Q_3$ and $t^c$, yielding two physical states with masses $m_t$ and~$m_T \simeq M \gg m_t$ (see experimental bounds on $m_T$ below). The mass eigenstates are linear combinations of the heavy and light tops, and the mixing angle for the left-handed components is given by
\beqa
\tan 2 \theta_L &\simeq& \frac{2 m_t }{ m_T } \frac{\lambda}{ y_t}  \, .
\eeqa
Denoting the physical mass eigenstates by $\tilde{t},\tilde{t}^c,\tilde{T}$ and $\tilde{T}^c$, we write the Yukawa Lagrangian in terms of these fields:
\beqa
\mathcal{L}_Y = &-& \cos^2 \theta_L \frac{m_t}{v_u}  h_u \tilde{t} \tilde{t}^c  -  \sin^2 \theta_L \frac{m_T}{v_u} h_u \tilde{T} \tilde{T}^c  \no \\
              &-& \sin 2 \theta_L  \frac{m_T}{2v_u} h_u \tilde{t}\tilde{T}^c - \sin  2 \theta_L  \frac{m_t}{2 v_u} h_u \tilde{T} \tilde{t}^c \no \\
              &+& {\rm h.c.}
\eeqa
The couplings of the light and heavy tops to the physical Higgs fields are given by:
\beqa\label{eq:VLQ}
c_{\tilde t}^i &=& \cos^2 \theta_L \left(c_t^i\right)_{\rm 2HDM} \, , \no \\
c_{\tilde T}^i &=& \sin^2 \theta_L \left(c_t^i\right)_{\rm 2HDM} \, ,
\eeqa
resulting in the following changes to the effective couplings to gluons and photons:
\beq\label{eq:cgVLQ}
\delta \hat{c}_g^i =\frac{9}{2}\delta \hat c^i_{\gamma}=  \sin^2 \theta_L \left[ A_f \left( \tau_T \right) - A_f \left( \tau_t \right) \right]\left(c_t^i\right)_{\rm 2HDM}\,.
\eeq

Existing bounds on vector-like top quarks give $M_T \gtrsim 475$~GeV, whether its primary decay channel is via charged-current~\cite{ATLAS:2012qe} or neutral-current~\cite{Chatrchyan:2011ay} interactions, or if the branching ratios in these channels are 1/2 and 1/4, respectively (as predicted by the equivalence principle in the large mass limit \cite{Kribs:2010ii}). We thus can safely use $A_f \left( \tau_T \right) \simeq 1$. Since the difference between the loop function of the light and heavy top particles is of order a percent, we find as expected that the results of a pure 2HDM are virtually unaffected by the presence of the new vector-like top.

Alternatively, one can add vector-like quarks in the representations $B \sim \left( 3,1 \right)_{-1/3}$ and $B^c \sim \left( \bar{3} ,1 \right)_{1/3 } $, which mix with the SM bottom quark. The couplings of the Higgs fields to the light and heavy mass eigenstates, $\tilde b$ and $\tilde B$, as well as the shift to the effective coupling to gluons and photons, can be straightforwardly deduced from Eqs.~\eqref{eq:VLQ} and~\eqref{eq:cgVLQ}. Experimental bounds on vector-like bottom quarks yield $M_B \gtrsim 480$~\cite{Chatrchyan:2012yea,CMS:2012jwa}, resulting again in $A_f \left( \tau_B \right) \simeq 1$.

The mixing with the electroweak-singlet $B$ alters the coupling of the $Z$ boson to the left-handed component of the light state $\tilde{b}$ by $\delta g_{b_L} = g_2\sin^2 \theta_L / (2\cos{\theta_W})$. This, in turn, changes the partial width of the $Z$ boson into two $b$ quarks:
\beq
\frac{\delta R_b }{R_b} \simeq 1.5 \frac{\delta g_{b_L }}{ g_{b_L}} \, .
\eeq
The experimental $2\sigma$ bound gives $-0.003\lsim\delta R_b/R_b\lsim 0.009 $~\cite{PDG}, leading to $\sin^2\theta_L\lsim~0.002$. Since the effective couplings to gluons and photons are proportional to $\sin^2 \theta_L$, they are highly suppressed under this constraint, and we find no deviation in the results from the pure 2HDM scenario.

%%%%%%%%%%
\mysection{2HDM plus a top-like scalar}
We now discuss the addition of one top-like scalar $\tilde t$ to the 2HDM scenario:
\beq
{\cal L} = {\cal L}_{\rm 2HDM}- \abs{\tilde{t}}^2 \left( M^2 + \lambda \abs{h_u}^2 \right)\,,
\eeq
which, in the language of~\eqref{eq:Leff}, with $S=\tilde t$, gives
\beqa
\tilde{m}_t^2 &=& M^2 + m_t^2 \frac{\lambda}{y_t^2}\,,\no\\
c_{\tilde t}^i &=& \frac{ m_t^2 }{\tilde{m}_t^2} \frac{\lambda }{y_t^2}\left(c^i_t\right)_{\rm 2HDM} \,,
\eeqa
resulting in a shift to the effective coupling to gluons and photons:
\beq\label{eq:stop}
\delta \hat c_g^i = \frac{9}{2}\, \delta \hat c_\gamma^i=\frac{1}{4} \frac{ m_t^2 }{\tilde{m}_t^2} \frac{\lambda }{y_t^2} A_s \left( \tau_{\tilde{t}} \right) \left(c^i_t\right)_{\rm 2HDM}\, .
\eeq
There are thus 2 additional parameters compared to the pure 2HDM case, $r\equiv\lambda/y_t^2$ and $\tilde m_t$. The quadratic divergences in the Higgs mass are canceled for $r=2$. We consider all values of $r$ up to the perturbative limit of $\pm4\pi$, and allow $\tilde m_t\gsim 80$~GeV. The pure 2HDM case is reproduced in the large $\tilde m_t$ limit. We find the following:
\begin{itemize}
  \item The two scalars $h$ and $H$ can provide for the conditions Eq.~\eqref{eq:cond} for positive $r$ and small $\tan\beta\lsim 0.6$ in type I for a variety of stop masses up to $\sim 1$~TeV. (For example, for $r=2$ the allowed mass range is $200-400$~GeV.) In this case, we find $R_{ZZ^*}\gsim 1.3$, $R_{\gamma\gamma}\gsim 1.7$, $R_{b\bar b}\sim 1.2$ and enhanced $R_{\tau\tau}\gsim 2.2$. In type II, larger $\tan\beta$ of order a few can be reached for low $\tilde m_t$, but the $\tau\tau$ rate is strongly enhanced to $R_{\tau\tau}\gsim 8$, excluding this scenario~\cite{ATLAStau}.
  \item If $\phi^1=h$ and $\phi^2=A$, the conditions of Eq.~\eqref{eq:cond} can be met in type~I for $\tan\beta\lsim 10$ for positive $r$ and $\tilde m_t\lsim 750$~GeV, but then  either $R_{\tau\tau}\gsim 2.5$ or $R_{b \bar b}\lsim 0.5$, disfavoring this scenario. The conditions Eq.~\eqref{eq:cond} are also met for $r<0$, but then $R_{b\bar b}$ is too low, disfavoring this scenario as well. In type~II, Eq.~\eqref{eq:cond} holds for $\tan\beta\lsim 0.7$ and positive~$r$, and combining the conditions of Eq.~\eqref{eq:condb} gives an enhanced diphton rate $R_{\gamma\gamma}\gsim 2$, suppressed diboson rate $R_{ZZ^*}\lsim 0.7$, $\tau\tau$ rate of $R_{\tau\tau}\sim 1.3$ and $R_{b\bar b}$ close to the SM value. For negative $r$, the allowed region yields a $b\bar b$ rate of order a percent. The conditions Eq.~\eqref{eq:cond} can be met in type~II for $\tan\beta$ of order a few as well, but then $R_{\tau\tau}\gsim 2.5$, disfavoring this scenario.
\end{itemize}

We learn that in a 2HDM with one additional top-like scalar, satisfying the conditions of Eq.~\eqref{eq:cond}, while keeping the $b\bar b$ and $\tau\tau$ rates at the reasonable levels of Eq.~\eqref{eq:condb}, is possible with the two scalars $h$ and $H$ in type I, or with $\phi^1=h$ and $\phi^2=A$ in type II, where both cases require small $\tan\beta$ below one. In the former case, enhanced rates are expected in the diboson, diphoton and $\tau\tau$ channels, with the $b\bar b$ channel proceeding slightly above the SM rate and $\tilde m_t\lsim 1$~TeV. In the latter case, the diphoton and $\tau\tau$ rates are enhanced, the diboson rate is suppressed and $R_{b\bar b}\sim 1$, with $\tilde m_t\lsim 700$~GeV. The results are summarized in Table~\ref{tab:sum}.

%%%%%%%%%%%%%%
\mysection{2HDM plus two top-like scalars}
We now move to the addition of two top-like scalars to the 2HDM. We denote by $\tilde t$ the up-component of an SU(2) doublet in the representation $(3,2)_{1/6}$ and $\tilde t^c\sim (\bar 3,1)_{-2/3}$. In the spirit of the MSSM, we consider the following stops-Higgs Lagrangian:
\beqa
\mathcal{L}_{\rm stop} = &-& \abs{\tilde{t}}^2 \left( \tilde{m}^2 + y_t^2 \abs{h_u}^2 \right) - \abs{\tilde{t}^c}^2 \left( \tilde{m}_c^2 + y_t^2 \abs{h_u}^2 \right) \no\\
&-& \left(y_t h_u X_t \tilde{t} \tilde{t}^c + {\rm~ h.c. }\right)+D{\rm-terms}\, ,
\eeqa
where in the MSSM $X_t = A_t - \mu \cot \beta$, and we do not write the $D$-terms explicitly. In the following we assume $X_t$ real. After EWSB, the two mass eigenstates, $\tilde{t}_1$ and $\tilde{t}_2$, are linear combinations of $\tilde{t}$ and $\tilde{t}^c$ with $m_1 \leq m_2 $ and mixing angle $0 \leq \theta_t \leq \pi$, given by
\beq
\begin{split}
\sin 2 \theta_t  = \frac{2 m_t X_t}{ m_2^2 - m_1^2 } \, .
\end{split}
\eeq

In the language of Eq.~\eqref{eq:Leff},
\beqa
c_{\tilde{t}_1}^i &=& \frac{m_t^2}{m_1^2} (c_t^i)_{\rm 2HDM} \left[ 1 - \frac{X_t}{2 m_t} \sin 2 \theta_t + D{\rm -terms}\right] ,\no\\
c_{\tilde{t}_2}^i &=& \frac{m_t^2}{m_2^2} (c_t^i)_{\rm 2HDM} \left[ 1 + \frac{X_t}{2 m_t} \sin 2 \theta_t + D{\rm -terms}\right],
\eeqa
where the $D$-term contributions above can be found {\it e.g.} in~\cite{Dermisek:2007fi}. The effective couplings to gluons and photons are changed according to:
\beq
\delta \hat{c}_g^i = \frac{9}{2}\delta \hat c^i_\gamma=\frac{1}{4} c_{\tilde{t}_1}^i A_{s}\left( \tau_{\tilde t_1} \right) + \frac{1}{4} c_{\tilde{t}_2}^i A_{s}\left( \tau_{\tilde t_2} \right) \, .
\eeq
In the limit of $m_2 \gg m_1 , m_t$ we find:
\beq\label{eq:2stop}
\delta \hat{c}_g^i \simeq \frac{1}{4} \frac{m_t^2}{m_1^2} (c_t^i)_{\rm 2HDM} A_{s}\left( \tau_1 \right)  \left[ 1 - \frac{X_t^2}{m_2^2} + 0.1 \times \cos 2 \beta \right] \, .
\eeq
The two top-like scalars contribution to the gluon effective coupling can be mapped to the one obtained in the single scalar case, Eq.~\eqref{eq:stop}. However, in the one-scalar language, only $r \lesssim 1.1$ is reached when adding two stops. We note that the inclusion of the $D$-terms does not qualitatively affect the results.

\begin{figure}[t]
\includegraphics[width=0.38\textwidth]{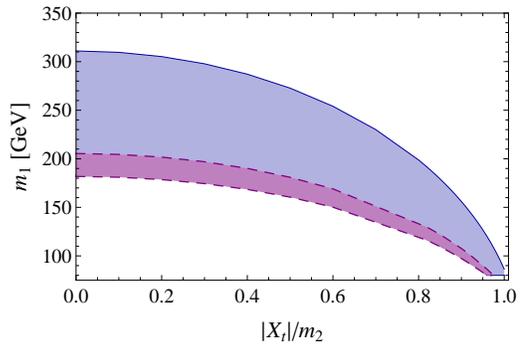}
\caption{Parameter space of a 2HDM with 2 top-like scalars in which Eqs.~\eqref{eq:cond} and~(3) hold, for $(\phi^1, \phi^2)= (h,H)$ type I (entire colored area) and $(h,A)$ type II (purple/lower).}
\label{fig:MSSM}
\end{figure}

We find, as expected, that the same scenarios are viable when adding one or two top-like scalars, and give similar rates in all considered channels, since there exists a mapping between the one- and two-stop cases.
In Fig.~\ref{fig:MSSM} we present the two-stops parameter space in which all the conditions of Eq.~\eqref{eq:cond} hold, and the $\tau\tau$ and $b\bar b$ rates obey Eq.~\eqref{eq:condb}. As is evident, the light stop should be lighter than $\sim 310$~GeV in all cases, and the mixing $X_t$ should obey $\abs{X_t}/m_2\lsim1$.

%%%%%%%%%%%%%%%%%%%%
\mysection{Conclusions}\label{sec:conc}
A Higgs-like boson of mass~$\sim125$~GeV has been discovered by the ATLAS and CMS collaborations, with significant observation in both the $\gamma\gamma$ and $ZZ^*\to 4\ell$ channels. In this work we posed the qualitative question of whether the signal seen in these two channels could be the result of two distinct resonances each contributing dominantly to a single channel. We investigated this idea in the context of a 2HDM as well as in several of its extensions. The (mild) conditions we use to quantify whether or not two separate particles can be at play in each channel are given in Eq.~\eqref{eq:cond} and include ratio (first two lines) and range (bottom two lines) conditions. In addition, we demand that the $b\bar b$ and $\tau\tau$ rates obey similar range conditions as the $ZZ^*$ and $\gamma\gamma$ channels, Eq.~\eqref{eq:condb}.

Table~\ref{tab:sum} summarizes our results of the viable scenarios: A 2HDM with one or two additional top-like scalars, where $(\phi^1,\phi^2)=(h,H)$ in type I or $(h,A)$ in type II. In each allowed case, we further detail the allowed range of $\tan\beta$, and indicate whether the diphoton, diboson, $b\bar b$ and $\tau\tau$ rates, as well as the dijet category in the diphoton channel, are enhanced, suppressed or similar to their SM value. We find that a ratio larger than three between the individual rates in the $\gamma\gamma$, $ZZ^*$ channels cannot be achieved subject to the range conditions.  All the allowed cases require low $\tan\beta$ below unity and an enhanced diphoton rate. The viable scenarios when adding one or two top-like scalars give similar rates in all considered channels, since there exists a mapping of one case to the other. The allowed two-stop parameter space is depicted in Fig.~\ref{fig:MSSM}.

\begin{table}[t]
\caption{The viable scenarios of the 2HDM in which the conditions Eq.~\eqref{eq:cond} and Eq.~\eqref{eq:condb} hold.
The scenarios with the replacement $h\leftrightarrow H$ are viable as well.}
\label{tab:sum}
\begin{center}
\begin{tabular}{|c|cc|c|c|c|c|c|c|c|} \hline
\rule{0pt}{1.2em}%
2HDM + & $\phi^1$ & $\phi^2$ & Type  & $\tan\beta$ & $R_{\gamma\gamma}$ & $R_{ZZ^*}$ & $R_{b\bar b}$ & $R_{\tau\tau}$ & $R_{\gamma\gamma}^{jj}$\cr \hline
1 or 2 stops & $h$ & $H$ & I & $\lsim 1$ & $\gsim 1.7$ & $\gsim 1.3$ & $\sim 1.2$ & $\gsim$ 2.2 & $\lsim 0.9$\cr
       & $h$ & $A$ & II & $\lsim 1$ & $\gsim 2$ & $\lsim 0.7$ & $\sim 1$ & $\sim 1.3$ & $\sim 1.5$\cr
\hline
\end{tabular}
\end{center}
\end{table}

As the Higgs data continues to accumulate, our knowledge of the rates in the $\gamma\gamma$, $ZZ^*$, $b\bar b$ and $\tau\tau$ channels, as well as in others, will improve significantly, further restricting the allowed cases and parameter space of the two-resonance scenario. Determining the parity of the boson in the $ZZ^*$ and $\gamma\gamma$ channels will also provide useful information in this context, as will improved constraints on masses of top-like scalars.

In the future, the two-resonance possibility studied in this work can result in two distinct outcomes. If the mass difference between the resonances is larger than the experimental resolution, with increased data two distinct resonances will be seen in the $ZZ^*$ and $\gamma\gamma$ channels, with a different resonance dominant in each channel. This is in contrast to a two resonance case where one particle is dominant in both channels~\cite{Gunion:2012gc,Ferreira:2012nv}. On the other hand, if the mass difference between the resonances is smaller than the experimental resolution, a double peak structure in the invariant mass distribution will not be detectable, and non-traditional methods, {\it e.g.}~\cite{Grossman:2013pt}, will need to be employed.

%%%%%%%%%%%%
\mysections{Acknowledgments}
We thank Liron Barak, Ofir Gabizon, Oram Gedalia, Eric Kuflik, Yotam Soreq, Michael Spira and Tomer Volansky for useful discussions, and are grateful to Yossi Nir for early collaboration, many helpful conversations and comments on the manuscript.

\end{document}